\documentclass[twocolumn,aps]{revtex4}
\usepackage{graphicx}
\usepackage{dcolumn}
\usepackage{bm}
\usepackage{amssymb}
\usepackage{amsmath}

\begin{document}

\title{Strain Effects on Point Defects and Chain-Oxygen Order-Disorder Transition in 123 Cuprate Compounds}

\author{Haibin Su}
\email{hbsu@wag.caltech.edu}
\altaffiliation{Present address: Beckman Institute 139-74, California Institute of Technology, Pasadena, CA 91125}
\affiliation{Department of Materials Science and Engineering, SUNY at Stony Brook, Stony Brook, NY 11794}
\author{David O. Welch}
\email{dwelch@bnl.gov}
\affiliation{Department of Materials Science, Brookhaven National Laboratory, Upton, NY 11973}
\author{Winnie Wong-Ng}
\email{winnie.wong-ng@nist.gov}
\affiliation{Ceramics Division, NIST, Gaithersburg, MD 20899}

\date{\today}

\begin{abstract}
\noindent

The energetics of Schottky defects in 123 cuprate superconductor series, $\rm
REBa_2Cu_3O_7$ (where RE = lanthandies) and $\rm YAE_2Cu_3O_7$ (AE = alkali-earths),
were found to have unusual relations if one considers only the volumetric strain. Our
calculations reveal the effect of non-uniform changes of interatomic distances within
the RE-123 structures, introduced by doping homovalent elements, on the Schottky
defect formation energy. The energy of formation of Frenkel Pair defects, which is an
elementary disordering event, in 123 compounds can be substantially altered under
both stress and chemical doping. Scaling the oxygen-oxygen short-range repulsive
parameter using the calculated formation energy of Frenkel pair defects, the
transition temperature between orthorhombic and tetragonal phases is computed by
quasi-chemical approximations (QCA). The theoretical results illustrate the same
trend as the experimental measurements in that the larger the ionic radius of RE, the
lower the orthorhombic/tetragonal phase transition temperature. This study provides
strong evidence of the strain effects on order-disorder transition due to oxygens in
the CuO chain sites.

\end{abstract}

\maketitle

\section{Introduction}

It is well known that during the fabrication process of superconductor materials, a
variety of point defects, such as substitution and interstitial impurities,
vacancies, and cation-disorder are involved. These defects have a large effect on the
properties of superconductors \cite{caizhu}. In type II superconductors, high
critical current densities can be achieved by the presence of high-density defects
which will provide suitable pinning centers for the magnetic flux lines. The ideal
size of defects for flux line pinning should be comparable to the superconducting
coherence length. For cuprates such as YBa$_2$Cu$_3$O$_{7-\delta}$ (Y-123), the
coherence lengths are in the order of tens of angstroms while the conventional
superconductors have a coherence length of several thousand angstroms. Thus
atomic-scale structural inhomogeneities such as point defects and columnar defects
can play an important role in flux-line pinning \cite{lieber96}. An increasing number
of applications of the 123-type high-$T_c$ superconductors use materials other than
Y-123. For example, in many bulk forms and multilayer applications, Y is replaced by
Nd, Sm, or other rare-earth elements (RE). Doping YBa$_2$Cu$_3$O$_{7-\delta}$ with
Ca, Sr, or other alkaline earth elements (AE) has also been shown to improve bulk and
grain boundary transport and other properties \cite{langhorn99, licci98, mannhart02}.

Since it is well-known that strain effects are important in the studies of point
defects, it is expected that studies of strain effects on point defects for series of
REBa$_2$Cu$_3$O$_{7-\delta}$ (RE-123) compounds will be important for practical
applications of superconductivity. In particular, the concentration and ordering of
oxygen vacancies have significant effects on the superconducting properties. For
Y-123, the superconducting temperature, $T_c$, depends on the oxygen stoichiometry.
As an example of a generic doping curve in cuprates \cite{Zhang93, Tallon95}, when
$\delta$ is larger than around 0.7, the crystal loses superconductivity. However, if
$\delta$ is smaller than 0.7, the compound is superconducting. It is generally
believed that higher oxygen content can create more holes in the structure. When
$\delta$ is between 0.1 and 0.7, the crystal is in the underdoped region, and $T_c$
increases with increasing hole concentration. When the oxygen content is in the
proper range, $T_c$ is above the boiling point of liquid nitrogen. However, if more
holes are created, the $T_c$ value decreases instead, and the crystal is in an
overdoped region.

Oxygen ordering in the Cu-O chains of the 123 structure gives rise to further complex
structures \cite{cava1, cava2}. Even when the average occupancies of oxygen sites
remain constant, the occupation at chain and anti-chain sites can vary. Jorgensen et
al. \cite{Jorgensen90} observed that the superconducting transition temperature in
Y-123 changes as a function of time following the quenching experiment while the
oxygen content is fixed. This demonstrates that the specific ordering of the oxygen
atoms in the basal plane is another important parameter that controls $T_c$ in the
Y-123 system \cite{Poulsen91}. The ordering process in the Cu-O chain is the origin
of the structural transition between tetragonal and orthorhombic phases in the 123
structure, which has been extensively studied since the discovery of Y-123
\cite{david87, deFontaine87, khachaturyan88, david89}. Structural transition of Y-123
is strongly affected by external pressure as well \cite{fietz96, veal97,
veal00,liar00}. In addition, there are systematic results reported by Wong-Ng et al.
\cite{winnie88, winnie89, winnie04} showing that the orthorhombic/tetragonal phase
transition temperatures in RE-123 can be scaled approximately linearly with the ionic
radius of RE$^{3+}$. The above experimental observations indicate that lattice strain
may play an important role in phase transition, which has not been investigated
theoretically and systematically so far.

In this paper, we plan to study the effects of homo-valent substitutions at Y and Ba
sites, and hydrostatic pressure on the order-disorder transition theoretically.
Considering large stress fields due to dislocations around grain boundaries, this
study will also provide valuable information for understanding the transport
properties in the vicinity of grain boundaries. First we briefly explain methods used
in atomistic simulations. Secondly, we focus on the effects of strain on Schottky
defects and related phenomena for a series of homovalent substitutions at Y and Ba
sites of YBa$_2$Cu$_3$O$_{7-\delta}$. Although the 123 structure is not stable in the
Ca- and Sr-analogs under ambient pressure \cite{roth89, okai90, davis91}, partial
substitution of Sr on the Ba site has been reported \cite{roth89, chu98a,
chu98b,licci98, licci00}. Finally, strain effects on chain-oxygen order-disorder
transition of REBa$_2$Cu$_3$O$_{7-\delta}$ will be investigated.

\section{Atomistic Simulation Methods}
\label{sec:methods}

Since the parent compounds of high-$T_c$ cuprate superconductors can be considered as
charge-transfer insulators, ionic bonding can be assumed to have a large contribution
to the lattice energy. Methods used to study atomistic phenonmena in conventional
oxides can therefore be applied to study cuprates. The cuprates become
superconductors at a proper doping level and temperature. However, the charge carrier
density is very low compared with that of conventional metals. In addition, the
charge carriers are confined within copper-oxygen planes. Consequently, the screening
effect is not as strong as that of conventional metals. The lattice energy calculated
from ionic models is somewhat over-estimated. The point defect's energy is also slightly overestimated due to the
omission of screening effect in metallic region of the phase diagram, which can be
improved by including polarization effects in the shell model \cite{2dick58}. Many
previous theoretical investigations are based on this type of ionic model (for
example,see \cite{2chaplot88, 2baetzold88, 2baetzold90, 2welch89, 2butler90}).

Several pair-potential sets of shell model parameters have been determined for Y-123
by Baetzold \cite{2baetzold88,2baetzold90}. To systematically study homo-valent
substitutions on Y- and Ba- sites, a consistent set of shell model parameters using
the data set for YBa$_2$Cu$_3$O$_7$ \cite{2baetzold88}was further developed to
account for the dependence of the Born repulsion of the two ions on their net
charges, on outer electronic configurations \cite{2pauling28, 2tosi64}, and on the
common ``$\rm r^3$ law" between polarizability and radius \cite{2bonin97}. The
``virtual crystal method" is applied here to interpret experimental data of mixing
two types of elements on one site. This method essentially is, for the purpose of
calculating average structural and elastic properties, to approximate the mixture of
two ions distributed over one sublattice by identical average ``virtual ions". It
allows the incorporation of compositional changes at the atomic level, but ignores
explicit effects of disorder. For instance, when Ba$^{2+}$ is partially replaced by
Sr$^{2+}$ in experiments, the composition-weighted average value of the two ions'
radii is taken to approximate that of each virtual divalent ion. After constructing a
consistent interatomic pair potential set \cite{su-thesis}, we used the
``General Utility Lattice Program" (GULP) \cite{gulp}, which integrates 
the above modeling methods at an atomistic scale, to study lattice 
energy, elastic constants, and lattice dynamic properties. In summary, our 
calculations are based on
short-range potentials of the Buckingham type, long-range Coulomb potentials, and
displacement-induced deformations of the electronic charge density in the framework
of a shell model.

\section{Schottky Defects in 123 Compounds}

\begin{figure}
\includegraphics [scale=0.2] {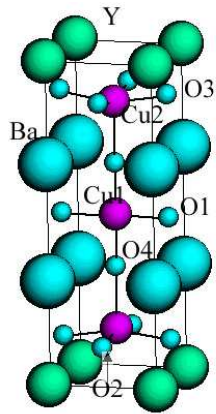}
\caption{ \label{fig:Y123Cell} Crystal structure of YBa$_2$Cu$_3$O$_7$.}
\end{figure}

\begin{figure}
\includegraphics [scale=0.4] {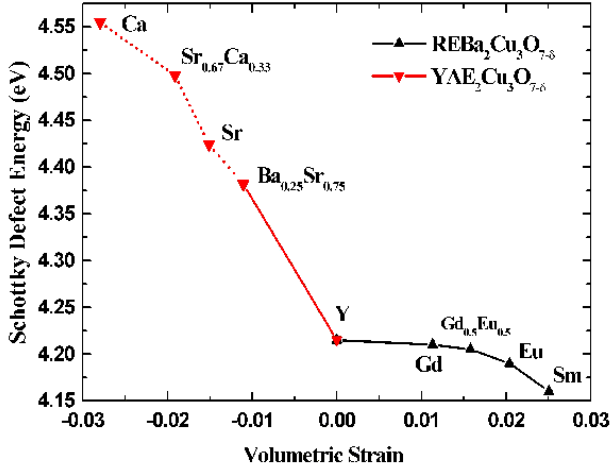}
\caption {\label{fig:Schottky} Schottky defects formation energy vs volumetric strain.
The volumetric strain is defined as ${V - V_0 \over V_0 }$, where $V_0$ is the volume
of Y-123 compound. All volumes are obtained by optimizing cell parameters and
internal coordinates to minimize the total energy. The solid solutions are treated by
the ``virtual crystal method".}
\end{figure}

\begin{figure}
\includegraphics [scale=0.4] {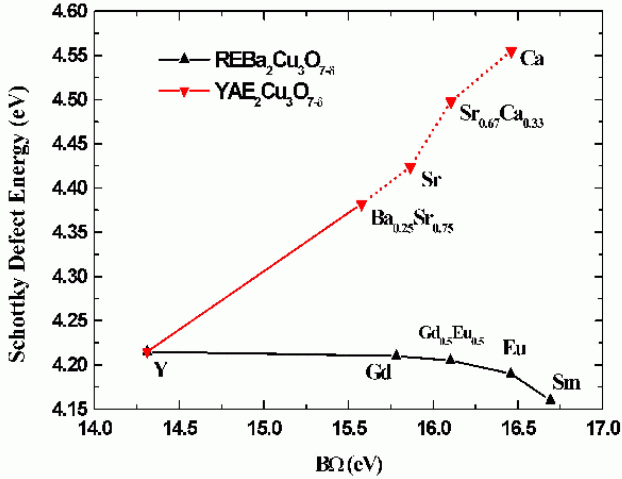}
\caption {\label{fig:varotsos} Schottky defects formation energy vs B$\Omega$ of 123
compounds. B is the bulk modulus and $\Omega$ is the mean volume per atom of 123
compounds. Both B and $\Omega$ are calculated from the optimized structures with
minimal total energy. The solid solutions are treated by the ``virtual crystal
method".}
\end{figure}

Any deviation in a crystal from a perfect structure is an imperfection. The simplest
imperfection is a lattice vacancy, which is a missing atom or ion, known as a
Schottky defect. Schottky defects involve ``multiple" vacancies while preserving
electrical neutrality. Regardless of how a Schottky defect is created, it is
necessary to expend a certain amount of work per atom to take it to the surface. We
calculated point defect energy by Mott-Littleton approach \cite{mott38, mott48}. Some
vacancies are at anionic sites and others at the cationic sites. For Y-213, the
defect reaction is given as follows:

\begin{equation}
YBa_2Cu_3O_7 \rightarrow V^{-3}_{Y} + 2 \cdot V^{-2}_{Ba}
+ 3 \cdot V^{-2}_{Cu} + 7 \cdot V^{+{13\over 7}}_{O} ;
\end{equation}

\noindent where V represents a vacancy. According to mass action law, the equilibrium
constant (K) at a finite temperature can be written as

\begin{equation}
\label{eq:mass}
K = c(V^{-3}_{Y}) \cdot  {c(V^{-2}_{Ba})}^2  \cdot  {c(V^{-2}_{Cu})}^3
\cdot {c(V^{+{13 \over 7}}_{O})}^7 = \bar{c}^{13};
\end{equation}

\noindent where c is the equilibrium concentration of vacancies; $\bar{c}$ is the
average concentration of vacancies. The equilibrium concentration of the vacancy j,
$c_j$, can be computed from Boltzmann statistics as

\begin{equation}
c_j = N_j \cdot \exp({-E_{v_j} \over kT });
\end{equation}

\noindent where $N_j$ is the number of atom j per unit volume, and $E_{v_j}$ is the
energy required to take atom j from its lattice site inside the crystal to a site on
the surface. Substituting the equilibrium concentraions of vacancies into equation
(\ref{eq:mass}), we obtain the expression for the average concentration of vacancies,
$\bar{c}$, as follows:

\begin{equation}
\bar{c} = N \cdot \exp({-\bar{E}_{Schottky} \over kT });
\end{equation}

\noindent where $N$ is the number of formula units per unit volume, and
$\bar{E}_{Schottky}$ is given, in terms of the formation energy of individual vacancy
(an ion is removed to infinity instead of the surface of crystals) and lattice energy
$E^{lattice}$, as

\begin{equation}
\bar{E}_{Schottky} = {\sum_{i} E_i^{vacancy} + E^{lattice} \over 13}.
\end{equation}

In RE-123, the absolute value of site potential decreases continuously with increasing
ion radius at Y-site, which is consistent with applying external tensile pressure.
Usually, the change of short-range repulsion is less significant than that of the
Madelung site potential. In general, the smaller absolute site potential value is
favorable for lowering vacancy energy, and vice versa. Furthermore, an increase of
the total volume leads to the expansion of effective relaxation space so that the
vacancy energy can further decrease due to the relaxation of the atoms. In the shell
model, the electronic polarization due to the electronic relaxation also lowers the
defect energy. In $\rm YAE_2Cu_3O_7$ (AE-123), as the radius of the ion at AE-sites
becomes larger, a similar phenomenon is observed. In both $\rm REBa_2Cu_3O_7$ and
$\rm YAE_2Cu_3O_7$, the lower Schottky defects formation energies are always
associated with the larger volume. This indicates that the volume of formation of
Schottky defects is positive, or the volume of the entire crystal expands during the
formation of Schottky defects.

As shown in Fig. (1), the complex structure of 123 compounds can be roughly divided into
six layers along c-axis direction: RE-CuO$_2$(Cu2)-BaO-CuO(Cu1)-BaO-CuO$_2$-RE. For
the sake of simplicity, the distances between planes are represented by the
separations of cations projected along c-axis. The external and internal strains are
computed with Y-123 as a reference. Replacing Y$^{3+}$ by larger RE$^{3+}$ ions leads
to positive external strains, indicating the dimensions of the cell expand as
increasing RE size. The more interesting observations are the changes of internal
parameters within the unit cell. The variance of internal strains provide direct
information of the changes of layers' separations in RE-series. The interatomic
distances vary quite differently with ionic radius in the RE-123 and AE-123 series
\cite{su-thesis}. For instance, the distance of RE-CuO$_2$ becomes larger
with increasing RE size, which is expected from the change of short-range repulsion
of RE-O. However, even when the entire cell volume increases as RE ion becomes
larger, some parts of unit cell contract such as interlayer separations between
CuO$_2$(Cu2)-BaO and BaO-CuO(Cu1) (or, bond lengths of Cu2-O4 and Cu1-O4).
Consequently, the average energy of the point defect only decreases slightly because
of the existing compressive blocks within the structure as plotted in Fig. (1). In the
AE-123 series, when Ba$^{2+}$ is (partially) replaced by smaller AE$^{2+}$, the
variances of external and internal strains are of the same sign, reflecting a
somewhat ``even" expansion of the entire structure. The difference between RE-123 and
AE-123 is reflected by the changes of AE-AE and Cu1-Cu2 distances. Unlike those in
RE-123, for AE-123 both distances become larger with increasing AE size. The
different changes of internal strains due to doping at either Y or Ba sites governs
the absolute value of slope of Schottky defects of RE-123 and AE-123 (see Fig.
(\ref{fig:Schottky})). In the previous studies \cite{islam89}, they found interesting correlations 
between dopant radius and energy of solution in YBa$_2$Cu$_3$O$_7$. The trend of subsituting divalent 
cation for barium matches well with that in the Schottky defect energy of YAE$_2$Cu$_3$O$_7$. However, 
there exist clear different trends between replacing yttrium by trivalent ions in YBa$_2$Cu$_3$O$_7$ and 
that in the Schottky defect energy of REBa$_2$Cu$_3$O$_7$. The reason is that the parent structure is fixed 
for former case, so that the trivalent ion larger than yttrium leads to more energy of solution. One the other hand,
the Schottky defect has to consider all the atoms in the unit cell. Unless the structures change more or less ``homogeneously", 
the energy of solution at one specific site may evolve differently from the Schottky defect energy.

Recall that the relation between cell volume, bulk modulus of RE-123, and the trend
of thermal expansion coefficients are also the manifestation of the complexity of the
123 structure \cite{su-thesis}. There is an ``unusual" relation between
Schottky defects formation energy and $B\Omega$ (where B is bulk modulus and $\Omega$
is the mean volume per atom) of RE-123 in Figure (\ref{fig:varotsos}). As reported by
Varotsos \cite{varotsos78}, the Schottky defects formation energy is proportional to
$B\Omega$ for elemental and binary crystals. While this relation appears to be obeyed
by the AE-123 series, it is violated by the RE-123 series. The possible reason is
still the non-uniform changes of the interatomic distances within the unit cell for
RE-123. We have studied the trend of melting temperature of RE-123 by ``Lindemann
law"  \cite{Gilvarry56, Poirier91}, and found that the vibrations along c-axis are
the most important modes to determine the melting temperatures while the isotropic
approximation fails to yield correct results \cite{su-thesis}. Here, we
take another approach to investigate the relation between melting temperatures and
formation energies of Schottky defects. Kurosawa \cite{Kurosawa56} gave a direct
correlation between them, which is listed in table (\ref{tab:DefectMelt}). While it
appears that a smaller Schottky defect formation energy can lower the melting
temperature of simple binary significantly, it has not been proved to be true for
compounds with complex structure. Note that the predominant point defect could be 
Frenkel defect rather than Schottky defect in some complex structures. However, Schottky defect 
energy provides an unambiguous measure on the average cohesive strength. For example, despite the Schottky 
defect formation energy of Sm-123 is smaller than that of Y-123, Sm-123 has a higher melting
temperature. It is necessary to study internal strains rather than volumetric strains
in order to obtain a detailed analysis of the thermodynamic properties of compounds
with complex structures. In general, we found that the signs of the slopes ($\rm
dE_{Schottky}/d\epsilon_v$) of RE-123 and AE-123 in Fig. (\ref{fig:Schottky}) are the
same. The difference between these two slopes and the ``unusual" relationships
between Schottky defects formation energies and $B\Omega$ and melting temperatures of
RE-123 compounds all reflect the complexity of the crystal structure.

\begin{table}
\begin{center}
\caption[The Schottky defects formation energy and melting temperature of binary and
RE-123 Compounds.] { The Schottky defects formation energy and melting temperature of
binary and RE-123 Compounds. The data of binary compounds are from Ref.
\cite{Kurosawa56}. The melting temperature data of RE-123 are from Ref.
\cite{5Osamura93}.} \label{tab:DefectMelt}
\begin{tabular}{rrrrrrr}
\hline
Crystal & NaF & NaCl & NaBr & MgO & CaO & SrO \\ \hline
$\bar{\rm E}_{\rm Schottky}(eV)$ & 2.5 & 2.2 & 2.0 & 6.3 & 5.5 & 5.0 \\
$T_{\rm melting}(^o\rm K)$ & 1259 & 1073 & 1018 & 3070 & 2850 & 2700 \\
\hline
\end{tabular}
\vspace{0.05in}
\begin{tabular}{rrrrr}
\hline
Crystal & YBCO & GdBCO & EuBCO & SmBCO \\ \hline
$\bar{\rm E}_{\rm Schottky}(eV)$ & 4.22 & 4.21 & 4.20 & 4.16 \\
$T_{\rm melting}(^o\rm K)$ & 1250 & 1290 & 1300 & 1325 \\
\hline
\end{tabular}
\end{center}
\end{table}

\section{Chain-Oxygen Order-Disorder Transition}

The oxygen content is an important parameter for superconductivity in cuprates. In
addition, the distribution (ordering) of oxygen atoms among the atomic sites strongly
affects $T_c$. The order-disorder transition of chain-oxygen has been extensively
studied both experimentally and theoretically \cite{cava1, cava2, david87,
deFontaine87, khachaturyan88, david89}. Raman studies show that there exists a
pressure-induced ordering phenomenon \cite{liar00}. The strain effect on this
transition has not been investigated theoretically in a systematic fashion. Here we
focus on the effects of homo-valent substitutions at Y and Ba sites, and hydrostatic
pressure on the order-disorder transition.

In the quasi-chemical approximation (QCA)\cite{david87, david89}, the short-range
order is characterized by the fraction number of near-neighbor pair sites occupied by
oxygen-oxygen pairs $p={N_{oo}\over 4N}$, where N is the number of sites on each of
the sublattice $\alpha$ and $\beta$, and $N_{oo}$ is the number of oxygen-oxygen near
neighbor pairs. The long-range order parameter S is defined such that the fractional
site occupancy on sublattice $\beta$ is $c(1+S)$, while that on sublattice $\alpha$
is $c(1-S)$ where c is the fractional site occupancy averaged over both sublattices.
[Note that for YBa$_2$Cu$_3$O$_{7-\delta}$, c is 0.5 when $\delta$ is zero; i.e.
$\delta$=1-2c.]

The partition function is given by

\begin{equation}
Z(T) = \sum_{R_{\alpha}} \sum_{q_{_{\alpha\alpha}}} g(R_{\alpha},
q_{_{\alpha\alpha}}
) e^{-W(R_{\alpha}, q_{_{\alpha\alpha}}) \over kT};
\end{equation}

\noindent where $R_{\alpha} = N(1+S)/4$; $q_{_{\alpha\alpha}}$ is the probability of
pairs $\alpha\alpha$; and $g(R_{\alpha}, q_{_{\alpha\alpha}})$ is proportional to the
total number of ways one can divide N entities into four groups of $\alpha\alpha,
\alpha\beta, \beta\alpha$, and $\beta\beta$ pairs. The configuration energy is
denoted as $W(R_{\alpha}, q_{_{\alpha\alpha}})$. As usual, we may replace each sum by
its maximum term in the summation for the system of a great many assemblies. Hence,
we have

\begin{equation}
Z(T) = \sum_{R_{\alpha}} g(R_{\alpha}, \bar{q}_{_{\alpha\alpha}})
e^{-W(R_{\alpha}, \bar{q}_{_{\alpha\alpha}}) \over kT};
\end{equation}

\noindent in which the most probable value $\bar{q}_{_{\alpha\alpha}}$ of
$q_{_{\alpha\alpha}}$ is determined by

\begin{equation}
{\partial \over \partial \bar{q}_{_{\alpha\alpha}}} (\ln g(R_{\alpha},
\bar{q}_{_{\alpha\alpha}})  - {W(R_{\alpha}, \bar{q}_{_{\alpha\alpha}})
\over kT} ) = 0 .
\end{equation}

There are three unknowns for a given value of temperature T and oxygen partial
pressure $P_{O_2}$: c, S, p. Three equations involving these three unknowns are
obtained by requiring that: 1) the chemical potential of oxygen atoms is the same on
both sublattices; 2) the chemical potential of oxygen atoms is the same in the solid
and in the gas phase(which consists mostly of diatomic molecules but has an
equilibrium concentration of atomic oxygen); and 3) the free energy of the system is
a minimum with respect to the fractional number of oxygen-oxygen pairs. The following derivations 
follow the similar procedure as described in \cite{muto55}. Using above
conditions, QCA yields

\begin{widetext}
\begin{eqnarray}
\ln ({c(1+S)-p \over c(1-S)-p}) & = & {3\over 4}
\ln ({(1+S)[1-c(1-S)] \over (1-S)[1-c(1+S)]}); \\
\ln [({1-c(1+S) \over c(1+S)})^3 &(& {c(1+S)-p \over 1-2c+p})^4]
= \ln [({P_{O_2} \over \xi})^{1\over 2} {1\over (kT)^{7\over 4}}]
- {\epsilon + {1\over 2 E_d} \over kT }; \\
{v\over kT} & = & \ln ({c(1+S)-p)(c(1-S)-p) \over p(1-2c+p)});
\end{eqnarray}
\end{widetext}

\noindent where $\epsilon$ is the energy to remove an oxygen atom from the gas and
place it in the lattice; $v$ is the repulsion energy between near-neighbour oxygens;
E$_d$ is the dissociation energy of one oxygen molecule; $\xi$ is equal to
4.144 $x$ 10$^{19}$ Pa(eV)$^{-7/2}$. The desired values of c, S, and p are obtained for
given values of T and $P_{O_2}$, by simultaneously solving the above equations. We
note that the order-disorder transition temperature T$_{od}$ is related to the
oxygen-oxygen repulsion energy $v$ and the value of the average site occupancy c at
this temperature by

\begin{equation}
\label{eq:vT} {v\over kT_{od}} = \ln ({16c(1-c) \over 1-4(1-2c)^2});
\end{equation}

\begin{figure}
\includegraphics [scale=0.4] {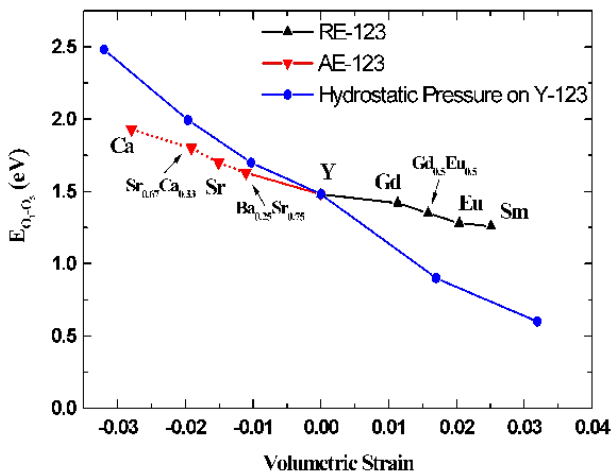}
\caption{\label{fig:Frenkel} Frenkel pair formation energy vs. volumetric strain for
123 Compounds. The energy corresponds to moving one chain-oxygen into an anti-chain
site. All volumes are obtained by optimizing cell parameters and internal coordinates
to minimize the total energy. The solid solutions are treated by the ``virtual
crystal method".}
\end{figure}

Frenkel defects involve an atom displaced from its normal site into an interstitial
site. If the interstitial site is chosen as anti-chain site for 123 compounds, this
Frenkel pair is closely related to chain-oxygen order-disorder transition. Forming a
Frenkel pair in an otherwise perfect crystal is an elementary disordering event. As
the disordering proceeds, it is important to account for defect-defect interactions.
Using Mott-Littleton approach, we have computed the isolated Frenkel pair formation
energy for REBa$_2$Cu$_3$O$_7$, YAE$_2$Cu$_3$O$_7$, and Y-123 under hydrostatic
pressure. We found that this formation energy increases in compressive regions and
decreases in tensile regions under hydrostatic pressure. This is the origin of
ordering under stress. From the systematic investigation of the phonon spectral
characteristics with the application of pressure \cite{liar00}, it was observed that
the changes induced by the hydrostatic pressure have a strong effect on  chain
ordering. Results of our calculations are consistent with the reported observation.
Figure (\ref{fig:Frenkel}) shows the plots of the Frenkel pair formation energy
versus lattice strain (volumetric strain) for both the RE-series and the AE-series.
These two plots demonstrate a similar trend and they form a well-connected smooth
curve. The similar behavior of the curve for the RE- and AE-series (with chemical
doping) and the curve for Y-123 (with hydrostatic pressure) indicates that the oxygen
disordering energy is dominated by lattice strain, which is expected if the short
range repulsion terms dominate the energy required to form the oxygen interstitial
ion.

\begin{figure}
\includegraphics [scale=0.4] {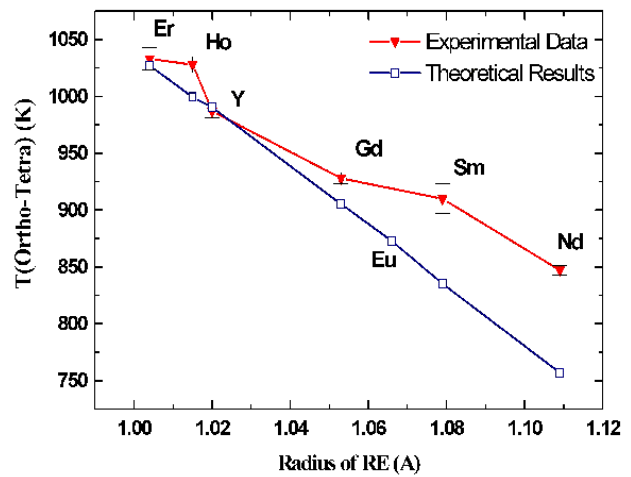}
\caption{\label{fig:Tod} Chain-oxygen order-disorder transition temperature of
REBa$_2$Cu$_3$O$_7$. The theoretical data are calculated by using the transition
temperature of $\rm YBa_2Cu_3O_{7-\delta}$ as a reference. The experimental data are
from Refs. \cite{winnie88, winnie89, winnie04}. The error bars indicate the
temperature range measured by X-ray Diffraction. }
\end{figure}

Furthermore, scaling the oxygen-oxygen short-range repulsive energy by the calculated
Frenkel Pair formation energy, the transition temperature between orthorhombic and
tetragonal phases is computed based on equation (\ref{eq:vT}). The transition
temperature of $\rm YBa_2Cu_3O_{7-\delta}$ was used as a reference. The results are
plotted in Fig. (\ref{fig:Tod}) with the experimental data taken from \cite{winnie88,
winnie89, winnie04}. There exists observable difference (c.a. 100 K) between theory
and experiment in Nd-123 system \cite{winnie04}. It has been determined
experimentally that the orthorhombic-to-tetragonal phase transition in the RE-123
series take place at an oxygen composition in the range of 6.4 (Er) to 6.83 (Nd), not
all at 6.5 (Y) \cite{winnie88, winnie89, winnie04}. The change of oxygen content
leads to the change of structure such as lattice parameters and atomic positions
\cite{cava1, cava2}, which in turn alters the formation energy of Frenkel pairs.
Since we set the reference transition temperature to be the value in Y-123 case, this
corresponds to the transition at oxygen content being 6.5. If we track the whole
process of order-disorder transition starting from fully oxygenated case, the total
oxygen content decreases until the transition is finished. The formation energy of
Frenkel pairs decreases also as anisotropy in the ab plane (defined as $\rm (b-a)/a$)
reduces. Note that the transition occurs at oxygen content being 6.83 in Nd-123
system, this indicates the anisotropy in Nd-123 remains higher compared with Y-123.
Therefore, the formation energy of Frenkel pairs decreases less than that in Y-123
during the whole process of transition as observed experimently. This means that in
our ``simple" model the transition temperature for Nd-123 is underestimated. But, the
remarkable thing is that using such a simple model, the trend of the theoretical
results agree well with the experimental measurements. It is seen that stress and
``chemical pressure" can substantially alter the degree of disorder. This study
provides a clear evidence for the effects of strain on order-disorder transition.
Previous studies also show there are quite rich microstructures resulting from this
orthorhombic-to-tetragonal phase transition: for instance twin structures and related twinning dislocations
\cite{chan1, chan2}, tweed morphology caused mainly by (110) and ($\bar{1}$10) shear
displacements \cite{caizhu}. In particularly, our results can be applied further to deduce that strain
near edge dislocations in low angle grain boundaries \cite{kung01}, which can also have a significant
effect on the degree of ordering in cuprate materials. For example, the interactions
between point defects and strain fields due to dislocations and/or grain boundaries
can affect the distribution of point defects, and the content and degree of order of
oxygen atoms \cite{su-thesis}.

\section{Conclusions}

Based on the Mott-Littleton approach we studied the Schottky defect formation energy
in the 123 phase as a function of volumetric strain. Generally, a more expanded
lattice favors a lower Schottky defect formation energy, and vice verse. The
difference of slopes ($\rm dE_{Schottky} / d \epsilon$) between RE-123 and
YAE$_2$Cu$_3$O$_{7-\delta}$, the ``unusual" relation between Schottky defects
formation energies and $B\Omega$ and melt temperatures of RE-123 compounds all
reflect the complexity of the crystal structure of 123 compounds.

Our study also illustrates the importance of strain effects on the
orthorhombic/tetragonal phase transition in the RE-123 compounds. We have calculated
the formation energy of Frenkel pair defects as a function of volumetric strain for
REBa$_2$Cu$_3$O$_7$ and YAE$_2$Cu$_3$O$_7$, and for YBa$_2$Cu$_3$O$_7$ under
hydrostatic pressure. Our calculations show good agreement with experimental
observations in that pressure favors ordering of the CuO chains. For example, the
Frenkel pair formation energy indeed increases significantly (c.a. -0.25 eV / 0.01
volumetric strain) under compression.  Based on a quasi-chemical approach, the
orthorhombic/tetragonal transition temperatures for RE-123 have been computed by
scaling the effective oxygen-oxygen short-range repulsive energy in the CuO chain
using the Frenkel pair formation energy. The calculated results agree with
experimental data in that the larger the ionic size of RE, the lower the
orthorhombic/tetragonal phase transition temperature.

\begin{acknowledgments}
HBS is grateful for Chinatsu Maeda's kind assistance in preparing the 
manuscript. The work at NIST was partially supported by the US Department 
of energy (DOE). The work at Brookhaven National Laboratory was performed 
under the auspices of the Division of Materials Sciences, Office of 
Science, U.S. Department of Energy under contract No. DE-AC-02-98CH10886.

\end{acknowledgments}

\end{document}